\documentclass[final,3p,times]{elsarticle}
\usepackage{times}

\usepackage[english,francais]{babel}
\usepackage{amsmath}
\usepackage{amsfonts}
\usepackage{amssymb}
\usepackage{amsthm}
\usepackage{color}
\usepackage{enumitem}
\usepackage{graphicx}
\usepackage{float}
\usepackage{array}
\usepackage{booktabs}
\usepackage{subcaption}
\graphicspath{{./plots/}}

\usepackage{bbm}
\usepackage{bm}
\usepackage{pstcol}
\usepackage{pstricks}
\usepackage{pst-text}
\usepackage{epsfig}
\usepackage{mathrsfs}
\usepackage{amscd}
\usepackage{xcolor}
\usepackage{makeidx}
\usepackage{slashbox}

\newcommand{\gT}{\bm{T}}

\newcommand{\gX}{\bm{X}}
\newcommand{\gY}{\bm{Y}}
\newcommand{\gZ}{\bm{Z}}

\newcommand{\eE}{\mathbb{E}}

\newcommand{\eR}{\mathbb{R}}

\newcommand{\sE}{\mathcal{E}}

\newcommand{\sH}{\mathcal{H}}

\newcommand{\sL}{\mathcal{L}}

\newcommand{\sN}{\mathcal{N}}

\newcommand{\sP}{\mathcal{P}}

\newcommand{\sX}{\mathcal{X}}

\newcommand{\sZ}{\mathcal{Z}}

\newcommand{\bfe}{{\textbf{e}}}

\newcommand{\gw}{{\textbf{w}}}
\newcommand{\gx}{{\textbf{x}}}

\newcommand{\gz}{{\textbf{z}}}

\newcommand{\galpha}{{\boldsymbol{\alpha}}}
\newcommand{\gbeta}{{\boldsymbol{\beta}}}

\newcommand{\gtheta}{{\boldsymbol{\theta}}}

\def\og{\leavevmode\raise.3ex\hbox{$\scriptscriptstyle\langle\!\langle$~}}
\def\fg{\leavevmode\raise.3ex\hbox{~$\!\scriptscriptstyle\,\rangle\!\rangle$}}

\usepackage[utf8]{inputenc}
\usepackage{glossaries}

\journal{arXiv}

\begin{document}

\begin{frontmatter}
\selectlanguage{english}
\title{From Risk Prediction to Risk Factors Interpretation. Comparison of Neural Networks and Classical Statistics for Dementia Prediction.}

\selectlanguage{english}

\author[PARIS]{Catherine Huber}
\ead{Catherine.huber@parisdescartes.fr}
\address[PARIS]{MAP5 CNRS 8145 @ University Paris Descartes \\
45 rue des Saints-Pères, F-75270 Paris Cedex 06, France.}

\medskip
\begin{center}
\end{center}

\begin{abstract}
\vskip 0.2\baselineskip
We are interested in predicting the onset of a disease D, based on several risk factors. For that purpose, two classes of techniques are available, whose properties are quite different in terms of interpretation, which is the focus of this paper :
\begin{enumerate}
\item Classical Statistics (for example: Generalized Linear Models (\textbf{GLM})).
\item Neural Networks (\textbf{NN}) (or more generally Artificial Intelligence (\textbf{AI})).
\end{enumerate}
Both methods are rather good at prediction, with a preference for Neural Network when the dimension of the potential predictors is high. But the advantage of the classical statistics is cognitive : the role of each factor is generally summarized in the value of a coefficient which is highly positive for a harmful factor, close to $0$ for an irrelevant factor, and highly negative for a beneficial one. While the underlying model in a neural network approach mixes repeatedly all factors together so that it is rather difficult to summarize the effect of each factor. However, we can reach some insight into interpretation of the respective impact of each risk factor using several algorithms 
In particular, we can distort the data set, doing sequential permutations of the risk factors. If the prediction performance of the neural network is stable, this means that the corresponding factor is irrelevant. Conversely, if the quality of the prediction decreases, the impact of the corresponding risk factor may be considered as proportional to this decrement.
\end{abstract}
\end{frontmatter}
\selectlanguage{english}
\section{Introduction}
Risk analysis is a topic of increasing importance in multiple fields like environment, technology,  medicine and biology. In survival data analysis and reliability, one is interested in risk factors that may accelerate or decelerate the life length of individuals or machines.
Some of them are endogenous (like genetic factors), others are exogenous (like pollution).
When analyzing the risk of an event $\sE$  to occur, such as a degradation, a failure, a disease or even death, one may consider how the waiting time $Y$ of onset of such a nocuous event is influenced by intrinsic and environmental factors $\gX:=(X_1,..,X_d)$:
\begin{equation}\label{general}
Y = f(\gX)
\end{equation}
The function $f$ is not deterministic. It has to be derived from the inspection of the potential risk factors observed on $n$ people, some who had the event and also some who did not experience the event. This leads to an estimation problem.\\
Or else, one may wonder whether the event $\sE$ occurs within a given amount of time $y_0$. In that case, one is faced with a discrimination problem: $Y'= 1 \{Y<y_0\}$ so that $Y'=1$ for people who experience the event within $y_0$ while $Y'=0$ for people who do not experience the event within $y_0$.\\
In classical survival data analysis, a stochastic model for $f$ is chosen among several families of models, fully parametric, nonparametric or semi-parametric. \cite{lee2019new,huber2004semiparametric,lawless2011statistical,huber2012goodness}. Let $\gx_i = (x_{i1},\cdots, x_{id})$ be the risk set observed on subject $i$, $i=1,\cdots,n,\, j=1,\cdots, d)$. Then, given a model $\sP$ and a loss function $\ell$, find $P$ in $\sP$ such that:
\begin{equation}\label{GLM}
\hat{L}(\hat{f})\;:=\;\min_{P\in \sP}\sum_{i=1}^n\frac{\ell(y_i,f(\gx_i))}{n}
\end{equation}
 The classical versions of these models are available in R software. To adapt the analysis to specific situations, researchers have to elaborate extensions of these models and work them out using R, which is both a software and a programming language; see \cite{pons2003estimation,therneau2013modeling}.\\
 As a counterpart, the machine learning approach of this same problem \cite{hastie2005elements,lecun2015deep} does not assume any model and
 leads to the so-called ``data driven models'', based on algorithms implying a set
 $\Theta$ of parameters in charge of minimizing the following expression :
 \begin{equation}\label{NN}
 \hat{L}(\widehat{N\!\!N_{\theta}})\;:=\;\min_{\theta\in \Theta}\sum_{i=1}^n\frac{\ell(y_i,N\!\!N_{\theta}(\gx_i))}{n}
 \end{equation}
 In that respect, it seems to be more satisfactory than the subjective choice of a stochastic model $\sP$ that appears in the first approach. However, machine learning is often viewed as a ``black box'' as the algorithm goes back and forth until convergence is achieved, and it scatters thus the initial potential risk factors in such a way that interpretation becomes difficult. However, every machine learning method, even though it seems to be purely algorithmic, has a probabilistic interpretation. We shall see this feature in particular for neural networks, which are a parametric version of a stochastic model: the projection pursuit regression and discrimination model.\\
 Now two important remarks should be mentioned.
\begin{enumerate}
\item Neural networks are no longer bounded to be black boxes :\\
  A NN is often considered as a black box between an entry $\gX$ and an output $Y$. However, the capacity of the algorithm of a NN to minimize the loss between the predicted value $\hat{Y}$ and the true value $Y$, (\ref{NN}), allows it to maximize the likelihood of a given probabilistic model $\sH$ including non linear functions. An example is a neural network extension of the Cox model in survival analysis. Also several possibilities were developed to interpret the role of each risk factor \cite{zhang2018opening}.
\item The problem of overparameterization
\begin{enumerate}
\item Overparameterization in classical statistics
\begin{itemize}
\item In the parametric setting, i.e. the model $\sP:=\sP_{\Theta}$ is defined up to a set of parameters $\theta \in\Theta \subset \eR^d$, increasing the number $d$ of parameters may lead to a perfect fit to the training set which may decrease the predictive ability on a new sample. For this reason, a penalization is applied, Lasso ($L^1$ norm) or  ridge ($L^2 $ norm) penalizations:
\quad\\\quad\\
Ridge regression shrinks the regression coefficients:
\[
\begin{array}{llllllllll}
\widehat{\theta}_{ridge} &=& \underset{\theta}{\arg\min}(\sum_{i=1}^n(y_i-\theta_0-\sum_{j=1}^p\theta_j\,x_{ij})^2 + \textbf{$\lambda\sum_{j=1}^d\theta_j^2$})
\end{array}
\]
\vspace{0.4cm}
\noindent Lasso regression also:
\vspace*{-7pt}
\[
\begin{array}{llllllllll}
\widehat{\theta}_{lasso} &=& \underset{\theta}{\arg\min}(\sum_{i=1}^n(y_i-\theta_0-\sum_{j=1}^d\theta_j\,x_{ij})^2 + \textbf{$\lambda\sum_{j=1}^d\vert \theta_j\vert$})\,\\
\end{array}
\]
\item In the non parametric setting, the penalization is done by a functional $J$ defined on $\sH$ so that what is to be minimized is a penalized empirical loss:
$$\min_{f\in \sH}\; [\hat{L}(f)\; + \lambda J(f)]$$
where
\[
\begin{array}{llllllllll}
\hat{L}(f)&:=&\sum_{i=1}^n\,\frac{\ell(y_i,f(x_i))}{n}\;\; \\
&&\\
J(f) &:=& \int_{\eR^d} \frac{|\tilde{f}(s)|^2}{\tilde{G}(s)}\,ds\\
&&\\
&&\tilde{f} \mbox{ : the Fourier transform of }$f$\\
&&\\
&&\tilde{G}(s)\mbox{ : a positive function such that }\tilde{G}(s)\xrightarrow[s \rightarrow \infty]{} 0.\\
\end{array}
\]
\quad\\
Example: space $\sH:=\sH_K$ of functions generated by a kernel $K$\\
Let $K(x,x')$ be a kernel, i.e. a continuous symmetric function of $x$ and $x'$ both in $\eR^d$, and of positive type i.e.:
$$\sum_{i=1}^n\sum_{j=1}^n K(x_i,x_j)c_ic_j \geq 0 \;\;\; \forall\; c_i, c_j \in \eR\;\; \mbox{ and } \forall\; x_i, x_j \in \eR^d$$
A basic example is the gaussian kernel:
$$K_h(x_0,x) = \frac{1}{h}\exp[\frac{-||x-x_0||^2}{2h}]$$
$\sH_K$ is the space spanned by linear combinations of $M$ functions $\{K(.,x_m), m = 1,\cdots, M\;\}$. Replacing $M$ by $n$ and $x_m$ by $x_i, i=1,\cdots,n$, the corresponding combination is an estimator of the probability distribution of $\gX$.
\end{itemize}

\item Overparameterization in neural networks\\
\noindent  Overparameterization in a neural networks approach seems to cause no problem (implicit or self-penalization?). It has been observed that, in deep learning, one can simultaneously
\begin{itemize}
\item fit perfectly the training set (empirical risk equals $0$),
\item have an efficient predictive ability on a new sample.
\end{itemize}
 In a recent paper \cite{bartlett2021deep}, the authors have a theoretical proof of this surprising phenomenon in a special case (p. 36-40, a two layers network) under certain conditions. We shall see that
\begin{itemize}
\item in our \textbf{simulation} study, a simple NN gets rid easily of the three irrelevant risk factors  $\textbf{Z}=(Z_1,Z_2,Z_3)$.
\item For the \textbf{real dataset}, predicting Alzheimer disease, NN is able, as well as GLM, to split the risk factors into two categories : the irrelevant and the relevant ones.
\end{itemize}
\end{enumerate}
\end{enumerate}


\section{Framework}
The purpose is to compare classical statistics to neural network approach for prediction of occurrence of a disease D both for prediction performance and interpretation of the risk factors impact.\\
This is done first on a simulation, then on a real data set of Alzheimer disease. The simulation
is based on a logistic model: a sample of size $n=1000$ with $d=6$ risk factors. Among them the first 3, $\textbf{X}=(X_1,X_2,X_3)$, are relevant,  defining the probability $p$ of occurrence of the disease,
 \begin{equation}
p:=P(Y=1\vert \gX=\gx):= \frac{\exp(a_1x_1+a_2x_2+a_3x_3)}{(1+\exp(a_1x_1+a_2x_2+a_3x_3))}
\end{equation}
The remaining factors $\textbf{Z}=(Z_1,Z_2,Z_3)$ are assumed to be irrelevant, i.e. independent of the outcome.\\
The real data set is a cohort of $n = 5003$ patients at Piti\'e Salp\'etri\^ere Hospital in Paris \cite{huber2019risk}. The expected prognostic is who will develop an Alzheimer within $y_0=4 $ years based on $d=13$ risk factors, including 3 genetic factors.\\
In both cases, we compare performance of a neural network and the classical logistic model.


\section{Neural networks}
A simple neural network has a single neurons layer and is a parametric version of a statistical semi-parametric process called Projection Pursuit Regression and Discrimination (PPRD):\\
\begin{enumerate}
\item Regression\\
The target $Y \in \eR$  is the response variable to $\gX=(X_1,\cdots,X_d) \in \eR^d$.
The PPR $\widehat{Y}$ of $Y$ is defined as:
\begin{equation}\label{def.PPR}
\widehat{Y} = \widehat{f}(\gX) := \sum_{m=1}^M \widehat{g_m}(\widehat{\gw_m^T} \gX):= \sum_{m=1}^M \widehat{g_m}(V_m)
\end{equation}
where $\gw_m, m=1,\cdots,M$ are unitary d-dimensional vectors and
$ g_m: \;\eR \rightarrow \eR$ ridge functions. Estimations are based on the observed training set: $(\gx_i, y_i), i=1,\cdots,n$. For $M$ big enough, any function can be approximated by (\ref{def.PPR}).
This is an additive model, but not with respect to the initial variables $\gX$ but with respect to
appropriate linear combinations of them: $V_m = \gw_m^T \gX$\\
Interpretation in terms of the initial inputs is difficult as each feature $X_j$ is scattered into every linear combination of $\gX$.
Usual error measurement is the quadratic error:
\begin{equation}\label{quadratic.error}
R(\gtheta):=\sum_{i=1}^n [y_i - \sum_{m=1}^M \widehat{g_m}(\widehat{\gw_m}^T\gx_i)]^2
\end{equation}
where $\gtheta$ is the set of parameters of the problem i.e. $\gw_m$ and $g_m$.\\
\item Discrimination: $K$ categories\\
For a discrimination problem, the response $Y$ is one of $K$ categories and the prediction $f_k(\gx_i)$ is the probability of category $k$ when $\gx=\gx_i$.  \\
Two error measurements are in use in that case:
\[
\begin{array}{llllllll}
R(\gtheta)&:=&\sum_{k=1}^K\sum_{i=1}^n (y_{ik}-f_k(x_i))^2 & \mbox{quadratic error}
&&\\
&&\\
R_{KL}(\gtheta)&:=-&\sum_{i=1}^n\sum_{k=1}^K y_{ik}\log(f_k(x_i))&\mbox{crossed entropy}
\end{array}
\]
\noindent The index $KL$ for the crossed entropy refers to Kullback Leibler ``distance" (not exactly a distance because lack of symmetry) of two probabilities $P$ and $Q$ which is defined as
\begin{equation}\label{Kullback}
KL(P,Q) = \int \log(\frac{dP}{dQ})\;dP
\end{equation}
\item Neural network as a special case of PPRD\\
 Our framework is a discrimination problem: the target $\gY=(Y_1,\cdots, Y_K)$ is a category, each $Y_k$ being a (0,1) variable to be predicted by $\gX=(X_1,\cdots, X_d)$.\\
$Y_k$ is modeled as a function $g_k$ of a linear combination of variables obtained
by a linear combination of {\em{activated}} $M$ linear combinations of the inputs.\\
A layer of $M$ neurons with entries $\gX$ produces a prediction $\widehat{\gY}$ of $\gY$ using
$(d+1)\times M$ coefficients $\alpha$ and $(M+1)\times K$ coefficients $\beta$.
Linearity comes in twice, with $(d+1)\times M$ coefficients $\alpha$
and $(d+1)\times K$ coefficients $\beta$.
\[
\begin{array}{llllll}
V_m&:=&\alpha_0 + \galpha_m^T X\;\;\;&m=1,\,2,\cdots,\,M\\
Z_m&=&\sigma(V_m)&\sigma\mbox{\em{ is the activation function}}\\
T_k &=& \beta_{0k} +\gbeta_k^T\gZ &k=1,\,2,\cdots,\,K\\
f_k(\gX)&=&g_k(\gT),& k=1,\,2,\cdots,\,K
\end{array}
\]
where $g_k(\gT) =\frac{e^{T_k}}{\sum_{i=1}^Ke^{T_i}}\;\Rightarrow \;$all $g_k(\gT)$ are positive and
add to 1.\\
$$\boxed{\widehat{Y_k}\,:=\,f_k(\gX)}$$ is the estimated probability of category $k$.

\item Minimize the error $R(\gY, \widehat{\gY})$ by an optimal choice of the parameters $\gw=(\galpha, \gbeta)$, obtained by gradient descent of $R$ with respect to $\gw$.
Possible choices for the activation function $\sigma$ are smoothed versions of the step function $s(u) = 1\,\{u \geq 0\}$:
\[
\begin{array}{llllll}
\sigma(u)&=&\displaystyle{\frac{1}{1+e^{-u}}}\;\;\;\mbox{the sigmoïd, the most usual one}\\
\quad\\
\sigma(u)&=&\displaystyle{\frac{e^u-e^{-u}}{e^u+e^{-u}}}\;\;\;\mbox{hyperbolic tangent (th(u))}\\

\sigma(a,u)&=&\begin{cases}
a(e^u - 1)\;\;\;&\mbox{for $u<0$}\\
u&\mbox{for $u\geq 0$}\mbox{ Exponential Linear Unit (ELU)}
\end{cases}\\

\sigma(a,u)&=&\begin{cases}
au\;\;\;&\mbox{for $u<0$}\\
u&\mbox{for $u \geq 0$} \mbox{ Rectified Linear Unit (ReLU)}
\end{cases}\\
\vspace{0.3cm}
\sigma(a,b,u)&=&b\begin{cases}
a(e^u - 1)&\mbox{for $u<0$}\\
u&\mbox{for $u \geq 0$ \mbox{Scaled Exponential Linear Unit (SELU)}}
\end{cases}
\end{array}
\]
\vspace{-0.1cm}
The nonlinearity of the model is due to the activation function. If $\sigma$
is the identity, the model becomes linear.
\begin{figure}[H]
	\centering
 \rotatebox{0}{\mbox{\resizebox{12.cm}{7.cm}			{\includegraphics[width=1\linewidth]{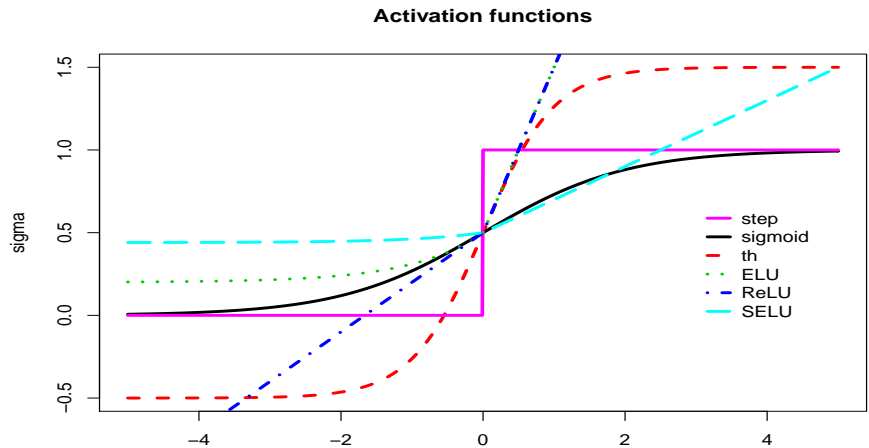}}}}
 \caption{Several activation functions}
\end{figure}
\end{enumerate}


\section{Comparing prediction and interpretation of GLM and NN on a simulation}
\subsection{The simulated data}
The simulated model is logistic:
\begin{equation}\label{logistic.model}
\ln(\frac{P(Y=1\vert (\gX,\gZ)=(x_1, x_2, x_3, z_1, z_2, z_3))}{P(Y=0\vert (\gX,\gZ)=(x_1, x_2, x_3, z_1, z_2, z_3))} = a_0 + a_1x_1 + a_2x_2 + a_3x_3 +  b_1z_1 + b_2z_2 + b_3 z_3 +\varepsilon
\end{equation}
where the relevant risk factors are $\gX=(X_1, X_2, X_3)$, $\varepsilon$ is a normal error, $\varepsilon \sim \sN(0,0.1)$ and
\begin{itemize}
\item $X_1$, binomial(p=0.3, size=3), coefficient $a_1=1$,
\item $X_2$, exponential(1), coefficient $a_2= 2$,
\item $X_3$, Poisson($\lambda=3$), coefficient $a_3=-1$.
\end{itemize}
\vspace*{0.2cm}
The irrelevant risk factors are $\gZ= (Z_1, Z_2, Z_3)$
\begin{itemize}
\item $Z_1$, binomial(p=$0.5$, size=$2$), coefficient $b_1=0$,
\item $Z_2$, normal($\mu=3,sd=1$), coefficient $b_2=0$,
\item $Z_3$, Poisson($\lambda=5$), coefficient $b_3=0$.\\
\end{itemize}
\begin{figure}[H]
	\centering
 \rotatebox{0}{\mbox{\resizebox{12.cm}{7.cm}{\includegraphics[width=1\linewidth]{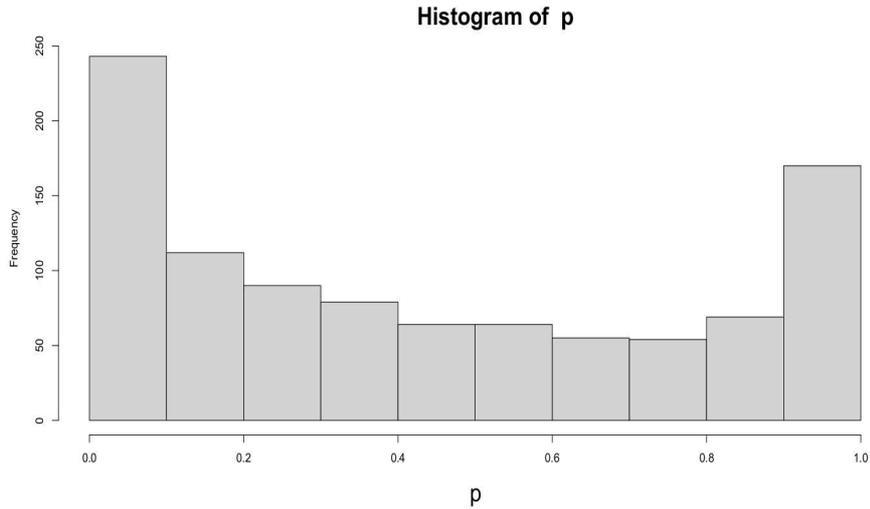}}}}
\caption{$yy:=a_1x_1+a_2x_2+a_3x_3+\varepsilon$,  $p =\frac{\exp(yy)}{1+\exp(yy)}$}
\end{figure}
\subsection{Prediction performances of GLM, the true model, and NN:}
The size of the training set is chosen to be equal to $2/3$ of the sample leaving $1/3$ for the test set. In the following table are presented the respective correct prediction probabilities for diseased ($p_d$), non diseased ($p_{nd}$) and global ($p_g$).
\begin{table}[H]
\begin{center}
\begin{tabular}{|c|c|c|c|c|c|}
\hline
 Method&$p_d$&$p_{nd}$&$p_g$&$CI_{95\%}(p_d)$&
 $CI_{95\%}(p_{nd})$\\
\hline
         GLM &0.833 &0.752& 0.788&0.827\;\;0.838&0.746\;\;0.758\\
\hline
        NN&0.857&0.752&0.808&0.849\;\;0.864&0.742\;\;0.762\\
\hline
\end{tabular}
{\center{\caption{Probability of correct predictions due to GLM and NN for diseased ($p_d$), non diseased ($p_{nd}$) global ($p_g$) and $95\%$ confidence intervals}}}
\end{center}
\end{table}
\vspace*{-0.2cm}
\subsection{Interpretation of risk factors impact by GLM and NN}
\begin{itemize}
\item GLM gives an estimation of the weight of every risk factor $\gx$ and $\gz$:
\begin{table}[H]
\begin{center}
\begin{tabular}{|c|c|c|c|}
\hline
        Risk factor&True coeff &coeff by GLM&p-value\\
\hline
        $x_1$&1&1.06 &$10^{-10}$\\
\hline
       $x_2$&2&2.04 &$5.5*10^{-27}$\\
\hline
        $x_3$&-1&-1.03&$5.110^{-28}$\\
\hline
        $z_1$&0 &-0.30& 0.23\\
\hline
        $z_2$&0 & 0.09&0.40\\
\hline
        $z_3$&0  &0.10& 0.050\\
\hline
\end{tabular}
\caption{Respective weights of risk factors $x_1, x_2, x_3$ (relevant) and $\gz$ (irrelevant) with corresponding p-values}
\end{center}
\end{table}
\vspace*{-0.5cm}
\item Neural Network (NN)\\
Before permuting every factor in turn, the mean probability to predict correctly D is $ p_d = 0.857 $.\\
After permutation of every factor in turn, the mean correct prediction becomes:
\begin{table}[H]
\begin{center}
\begin{tabular}{|c|c|c|c|}
\hline
        m.x1&m.x2 &m.x3&relevant factors\\
\hline
        0.842& 0.762& 0.787& $< 0.857$\\
\hline
        m.z1&  m.z2&  m.z3&irrelevant factors\\
\hline
       0.857& 0.856& 0.855& $\approx$ 0.857\\
\hline
\end{tabular}
\caption{Mean correct probability of prediction of occurrence of the disease $p_d$ when doing N=100 permutations of each risk factor $x_1, x_2, x_3, z_1, z_2, z_3$. }
\end{center}
\end{table}
\vspace*{-0.5cm}
\hspace*{-1cm}Conclusion:\\
\hspace*{-0.5cm}$z_i$'s permutation does not change the probability of a correct prediction.\\ \hspace*{-0.5cm}$x_i$'s permutation reduces the probability of a correct prediction, with a predominance of the impact of $X_2$ which can be seen also in the classical statistic approach.
\end{itemize}


\section{Comparing prediction and interpretation of NN and GLM on Alzheimer data:}
\subsection{Description of the data set}
A cohort of $5003$ patients was collected at Piti\'e Salp\'etri\^ere Hospital in Paris in order to study the onset of Alzheimer. The final sample, after verification, has $n=4356$ patients. The risk factors considered in the sample were age at inclusion, gender, education, cardiac disease, depress, incapacity, high blood pressure, birth date, three genetic factors (APOE4, $\cdots$ ). Among them, $n_1=142$ developed an Alzheimer within $4$ years. The issue was to predict who will develop an Alzheimer knowing his, or her risk factors. \\
We compared neural network (NN) with a classical logistic model (GLM) in this setting, where $Y$ is equal to $1$ for patients who became Alzheimer within 4 years and $0$ otherwise.
\begin{equation}\label{logistic.Alz}
    P(Y=1\vert \gX=\gx) = \frac{\exp(\gw^T\gx)}{1+\exp(\gw^T\gx)}
\end{equation}
Note that the very unbalanced counts for diseased ($142$ for $Y=1$) and controls ($4214$ for $Y=0$) creates difficulties for prediction which can be overcome as we shall see.
\vspace{-0.3cm}
\subsection{Prediction performances of GLM and NN for Alzheimer:}
\begin{enumerate}
\item First, split at random $3/4$ of the data set to be the training set. The remnant ($1/4$) will be the test set, on which to predict who will be Alzheimer. Use separately logistic model (GLM) and neural network (NN) on the training set to estimate the probabilities $p_d(x)$ to develop a dementia (Alzheimer) based on the risk factors $\gX=x$. Then predict, on the test set, who will be Alzheimer based on the estimations done with both methods.
    \quad\\The result is four counts for each method:
    \begin{itemize}
    \item true positive,
    \item false positive,
    \item true negative,
    \item false negative.
    \end{itemize}
\item Repeat this process $N$ times, for both methods, to obtain confidence intervals for the probability of correct prediction.
\end{enumerate}
\vspace*{-0.2cm}
\begin{table}[H]
\begin{center}
\begin{tabular}{|c|c|c|c|c|c|}
\hline
 Method&$p_d$&$p_{nd}$&$p_g$&$CI_{95\%}(p_d)$&
 $CI_{95\%}(p_{nd})$\\
\hline
         GLM &0.72 &0.73& 0.73&0.55\;\;0.85&0.70\;\;0.76\\
\hline
        NN&0.68&0.73&0.73&0.50\;\;0.85&0.65\;\;0.77\\
\hline
\end{tabular}
\caption{Correct predictions due to GLM and NN for dements ($p_d$), for non dements ($p_{nd}$), global $p_g$ and $95\%$ confidence intervals}
\end{center}
\end{table}
\vspace*{-0.4cm}
\noindent Some comments
\begin{enumerate}
\item The fact that the counts are very much unbalanced ($142$ dements versus more than $4000$ non dements) creates problems for the prediction: the confidence intervals are large.
\item To overcome this problem, one can duplicate the smaller category \cite{lecun2015deep,lecun2018perso}.
\begin{table}[H]
\begin{center}
\begin{tabular}{|c|c|c|c|c|c|c|c|}
\hline
 Method&$p_d$&$p_{nd}$&$p_g$&$CI_{95\%}(p_{d})$&
 $CI_{95\%}(p_{nd})$)\\
\hline
         GLM &0.73 &0.73& 0.73&0.71\;\;0.76&0.71\;\;0.75\\
\hline
        NN&0.75&0.72&0.73&0.73\;\;0.78&0.70\;\;0.75\\
\hline
\end{tabular}
\caption{ Correct predictions due to GLM and NN for dements ($p_d$), for non dements ($p_{nd}$), global $p_g$, and $95\%$ confidence intervals after duplication}
\end{center}
\end{table}
\vspace*{-0.5cm}
\item The widths of the $95\%$ confidence intervals are reduced
$[0.71\;\;0.76]$ instead of $[0.50\;\;0.85]$ for the future Alzheimer detection
$[0.70\;\;0.75]$ instead of $[0.65\;\;0.77]$ for the future non Alzheimer\\
\end{enumerate}
\vspace*{-0.2cm}
\subsection{Interpretation for GLM and NN}
\begin{itemize}
\item GLM\\
Interpretation is much easier in classical statistics. Respective influence of the risk factors are available from the probabilistic modelling. Weights of the risk factors obtained by the logistic model:\\
age is compared to age  $< 70$
\[
\begin{array}{lllllllllll}
\mbox{Age } \in [70\;80] &:&\mbox{ risk multiplied by }&3&( 3.1, CI_{95\%} =[1.6\; 5.9]\\
\mbox{Age }>80 &:& \phantom{ risk multiplied by }  &8&(8.3, CI_{95\%} =[4.3 \;16]\\
\mbox{Cardiac disease }&:& \phantom{ risk multiplied by } &2& ( 1.9, CI_{95\%} =[1.2\; 2.9]\\
\mbox{Depress }&:&\phantom{ risk multiplied by }  &2.5&( 2.3, CI_{95\%} =[1.5\; 3.3]\\
\mbox{Incapacity }&:&\phantom{ risk multiplied by } &3.5&( 3.4, CI_{95\%} =[2.2\; 5.1]\\
\mbox{APOE4 }&:& \phantom{ risk multiplied by }  &2&( 1.9, CI_{95\%} =[1.3\; 2.8]
\end{array}
\]
This motivates the reluctance of certain statisticians to use Machine Learning. But it is nowadays changing rather fast.
\item NN: Risk factors impact for Neural Networks
\begin{table}[H]
\begin{center}
\begin{tabular}{|c|c|c|c|c|c|c|c|c|}
\hline
 Permutation&$p_d$&$p_{nd}$&$p_g$&$CI_{95\%}(p_{d})$&$CI_{95\%}(p_{nd})$)&\\
\hline
none &0.7553&0.7739&0.7650 &0.7412\;\;0.7694&0.7662\;\;0.7758&\\
\hline
AA&0.7419&0.7724&0.7581&0.7395\;\;0.7442&0.7699\;\;0.7749&$\approx$\\
\hline
AG&0.7457&0.7751&0.7613&0.7418\;\;0.7495&0.7717\;\;0.7786&$\approx$\\
\hline
age&0.7098&0.7410&0.7264&0.7057\;\;0.7139&0.7338\;\;0.7481&$\downarrow$\\
\hline
APOE4&0.7341&0.7629&0.7494&0.7289\;\;0.7393&0.7594\;\;0.7665&$\downarrow$\\
\hline
card&0.7446&0.7748&0.7606&0.7401\;\;0.7491&0.7721\;\;0.7775&$\approx$\\
\hline
CC&0.7473&0.7779&0.7635&0.7428\;\;0.7518&0.7747\;\;0.7811&$\approx$\\
\hline
depress&0.7381&0.7671&0.7535&0.7343\;\;0.7420&0.7636\;\;0.7706&$\downarrow$\\
\hline
education&0.7473&0.7772&0.7632&0.7444\;\;0.7503&0.7748\;\;0.7797&$\approx$\\
\hline
gender&0.7447&0.7758&0.7612&0.7403\;\;0.7490&0.7725\;\;0.7792&$\approx$\\
\hline
HTA&0.7510&0.7808&0.7668&0.7457\;\;0.7564&0.7765\;\;0.7852&$\approx$\\
\hline
incapacity&0.7282&0.7609&0.7455&0.7243\;\;0.7320& 0.7584\;\;0.7634&$\downarrow$\\
\hline
psy&0.7419&0.7724&0.7581&0.7395\;\;0.7442&0.7699\;\;0.7749&$\approx$\\
\hline
TC&0.7465&0.7773&0.7628&0.7450\;\;0.7480&0.7748\;\;0.7799&$\approx$\\
\hline
\end{tabular}
\caption{ Effect, on prediction ability, of permutation of each risk factor
AA, AG, CC, TC are genetic factors like APOE4}
\end{center}
\end{table}
\vspace*{-0.3cm}
We see in the last column of this table that the probability of correct prediction decreases for age, gene APOE4, depress and incapacity, while it remains rather stable when permuting the values of every other factor. Except for the cardiac disease, for which it is not clear, NN and GLM have matching results on relevant factors leading to Alzheimer, age and incapacity being the strongest in both cases.
\end{itemize}


\begin{figure}[H]
	\centering
 \rotatebox{0}{\mbox{\resizebox{12.cm}{7.cm}			{\includegraphics[width=1\linewidth]{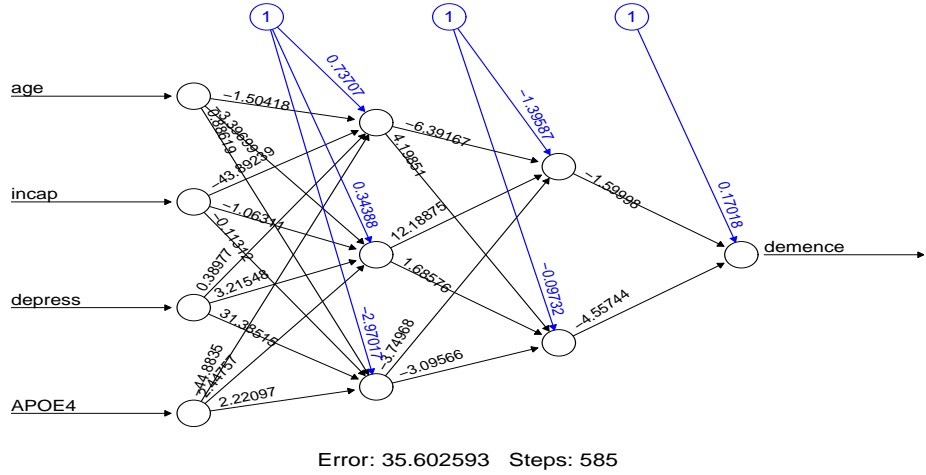}}}}
\caption{2 layers, 3 and 2 neurons}
\end{figure}

\section{Conclusions and perspectives}
\begin{enumerate}
\item We have seen that, in the special case of a moderate number of risk factors, the respective prediction performances of a probabilistic model (logistic) and a neural network were similar, both on a simulation and on real data of Alzheimer patients. Moreover, we have seen that the interpretation of the impact of each risk factor is also similar though easier and more precise for the probabilistic model.\\
This is not really surprising as a neural network approach, in its basic version, is a parametric version of a non parametric statistical model namely Projection Pursuit Regression and Discrimination model (PPRD). In both cases, interpretation in terms of the risk factors is difficult as the risk factors are scattered and mixed along the solving algorithm.\\
Also, one must notice that it is possible to use a NN approach to solve a probabilistic model. For example, the Cox model \cite{cox2018analysis}, which is the preferred model to analyze survival data, can be solved by a NN approach \cite{katzman2018deepsurv}.\\
The maximization of the Cox partial likelihood, $\sL_c$, can be obtained through a NN, which  minimizes a function analog to $-\sL_c$, replacing the linear function $\gw^T \gx$ by a nonlinear one $h_{\theta}(\gx)$:
    \begin{equation}
    \sL_c(\gw)=\prod_{i}\delta_i\frac{\bfe\,^{\gw^T \gx_i}}{\sum_{j:t_j\geq t_i}\bfe\,^{\gw^T}\gx_j}
   \end{equation}
    \quad\\
   \begin{equation}
    \sL_{NN}(\theta)=-\,\prod_{i}\delta_i\frac{\bfe\,^{h_{\theta}(\gx_i)}}{\sum_{j:t_j\geq t_i}\bfe\,^{h_{\theta}(\gx_j)}}
    \end{equation}

    \quad\\
where $\delta_i = 1$ if subject $i$ failed at time $t_i$, and $0$ if subject $i$ is censored at time $t_i$.\\
    The loss function minimized by the NN with parameters $\theta$ is $-\sL_{NN}(\theta)$.
\quad\\
\quad\\
The nonlinearity of the NN approach, due to the activation function $\sigma$, allows to approach any function as close as wished.\\

\item In this paper, we were faced with a moderate number of risk factors, which is rather favorable for probabilistic models.\\
It is thus necessary to have the same kind of comparison both for prediction and interpretation when the number of the risk factors is huge. In that case, we would need at least two preliminary procedures:
\begin{itemize}
\item For the probabilistic model, use a preliminary method to reduce the dimension. Among the numerous devices whose purpose is to reduce the dimension of the entries like PCA (Principal Component Analysis, SVD (Singular Value decomposition), MDS (MultiDimensional Scaling) most of them are linear. \\
However, based on the $K$ nearest neighbours $(j_1, j_2,\cdots, j_K)$ of every point $i$ in the input space $\sX$ assumed to be a metric space ($\eR^d$ in general), a weighted graph is built, the weight of each edge $(i,j_k)$ being equal to $d(i,j_k)$, and a geodesic distance.
The {\em{geodesic distance}} of any pair of points $(i,j)$ in the graph is the minimum path between them. This leads to discover the structure of the data, which may be a manifold rather than a linear subspace as is the case in PCA, SVD and also MDS.
\item For the neural network approach, the problem is not relative to the prediction as NN eliminates easily the irrelevant factors. The problem is rather due to the proposed method to estimate the impact of each risk factor: permuting each factor in turn would be time consuming so that one should try to randomly permute random subsets of risk factors.
\end{itemize}
\item Our method should be compared to the existing methods which are numerous:
\begin{itemize}
\item Garson's algorithm for a one layer neural network \cite{garson1991comparison}:\\
The weights connecting neurons to a NN are partially analogous to the coefficients in a GLM model. Garson's algorithms relies upon he fact that the combined effects of the weights on the prediction represent the relative importance of each predictor.
The formula that gives the relative importance of entry $x_{\ell}$ may be read as
\vspace*{0.5cm}
\begin{equation}
\frac{\sum_{j=1}^{n_H}(\frac{w(x_\ell,j)}{\sum_{i=1}^{d}w(x_i,j)}w(y,j))}
{\sum_{i=1}^d(\sum_{j=1}^{n_H}(\frac{w(x_{\ell},j)}{\sum_{i=1}^{d}w(x_i,j)}w(y,j)))}
\end{equation}
\quad\\
\quad\\
where $n_H$ is the number of hidden nodes, $d$ the dimension of the input $\gx$, $w(x_i,j)$ the weight of entry $x_i$ at node $j$, and $w(y,j)$ the output weight at node $j$.
\item Lek's profile method \cite{lek1996application}\\
This method may be applied only when the predictors are continuous.\\
As the relationship between a predictor and an outcome may depend on the values of the other predictors, Lek's profile explore the relationship between a chosen predictor while holding other predictors in a set of constant values (e.g. minimum, some quantile, maximum). The method generates a partial derivative of the response with respect to each predictor.
\item Shapley value to measure importance of dependent inputs \cite{owen2017shapley}. The value of a subset $u$ of $\{1, 2, \cdots, d\}$ is the explanatory power of $x_u$:
\begin{equation}
val(u) := var(\eE(f(x)\vert x_u ))
\end{equation}
As desirable properties for the attribution of a value $\phi_i$ to every entry $x_i$ are
\begin{itemize}
\item $\sum_{i=1}^d \;\phi_i \;= \; val(1:d)$
\item If $val(u \cup \{i\}) = val(u)\;\; \forall u \subseteq (1:d)$, then $\phi_i=0$.
\item If $val(u\cup \{i\})= val(u\cup \{j\})\;\; \forall u \subseteq (1:d)-\{i,j\}$, then $\phi_j=\phi_i$.
\item If $val$ and $val^*$ have Shapley values $\phi$ and $\phi^*$ respectively, then the "game"    with value $val$ + $val^*$ has Shapley values $\phi_i + \phi^*_i$.
\end{itemize}
the only valuation $\phi$ that meets those axioms was proved to be the following \cite{shapley1953stochastic}
\begin{equation}
\phi_i = \frac{1}{d} \sum_{u \subseteq (1:d)-\{i\}} \frac{val(u \cup \{i\})-val(u)}{C_{d-1}^{|u|}}
\end{equation}
where $C_n^k$ is the number of combinations of n by k.
\item LIME method: Local Interpretable Model-agnostic Explanations\cite{pedersen2018lime}.\\
 This method is based on approximating the NN, $f(x)$, locally (in the vicinity of $x$) by an interpretable model $g \,\in \, G$, where $G$ is a class of interpretable (simplified) model whose complexity is defined as $\Omega(g)$. The simplified model $g$ is a function of {\em{interpretable}} representations $x'$ of the initial features $x$. While $x \in \eR^d$, $x'$ may be a binary vector $x' \in \{0,1\}^{d'}$, $d'<d$.\\
 This allows the approximate model to change when the neighborhood of the explanatory variables $x$ changes, which happens when the relationships between inputs and outputs are non linear. To define a vicinity of $x$, let $\pi_x(z)$ be a proximity measure between $z$ and $x$.\\
 In the classification setting, $f(x)$ is the probability (or  binary indicator) that $x$ pertains to a certain class.\\
 In classical statistics, the trade-off is between bias and variance, while here the trade-off is between local {\em{Fidelity}} and global {\em{Interpretability}}. As the interpretability is a decreasing function of the complexity $\Omega(g)$ of model $g$ and local fidelity is a decreasing function of $\sL(f, g, \pi_x)$ defined as a measure of how unfaithful $g$ is in approximating $f$ in the locality defined by $\pi_x$, the explanation produced by LIME is obtained by the following equation
 \begin{equation}\label{lime}
\xi(x) = \arg\min(\sL(f, g, \pi_x) + \Omega(g))
 \end{equation}
Different explanations result from the choice of the three quantities, $G$ the family of interpretable models, fidelity functions $\sL(f, g, \pi_x)$ and complexity $\Omega(g)$.\\
An approximation of $\sL(f, g, \pi_x)$ for some chosen $x$ is obtained by drawing samples weighted by $\pi_x$. Given a perturbation $z'\in \{0,1\}^{d'}$, which contains a fraction of the non-zero elements of $x'$ associated to $x$, the interpretable representation of $x$, we recover the sample in the original representation $z \in \eR^d$ and obtain $f(z)$, which is used as a {\em{label}} for the explanation model. Given this dataset $\sZ$ of perturbed sample with the associated labels, one optimizes (\ref{lime}) to get explanation $\xi(x)$.
\item Recent approaches to interpretation of NN aim at unifying local and global explainability \cite{lundberg2017unified,GIUDICI2021114104,giudici2022explainable}.
\end{itemize}
\end{enumerate}


\end{document}